\documentclass[aps,pra,preprint,groupedaddress,showpacs]{revtex4}
\usepackage[latin1]{inputenc}
\usepackage{eurosym}
\usepackage{makeidx}
\usepackage{graphicx}
\usepackage{amsfonts}
\usepackage{fancyhdr}
\usepackage{setspace}
\usepackage{amssymb}
\usepackage{anysize}

\begin{document}

\title{Time dependence of quantum entanglement in the collision of two particles}

\author{Mihály G. Benedict} \email{benedict@physx.u-szeged.hu} 
\author{Judit Kovács} 
\author{Attila Czirják} \email{czirjak@physx.u-szeged.hu}
\affiliation{Department of Theoretical Physics, University of Szeged, 
\\
H-6720 Szeged, Tisza Lajos körút 84-86, Hungary}

\begin{abstract}
We follow the emergence of quantum entanglement in a scattering event between
two initially uncorrelated distinguishable quantum particles interacting via a
delta potential. We calculate the time dependence of the Neumann entropy of
the one-particle reduced density matrix. 
By using the exact propagator for the delta potential, we derive an approximate analytic
formula for the asymptotic form of the two-particle wave function which is
sufficiently accurate to account for the entanglement features of the system.
\end{abstract}




\maketitle

\section{Introduction}

Quantum entanglement has become an important research topic in modern physics,
not only because it exhibits the striking differences from classical concepts,
but also since it is widely considered now as the fundamental resource in
quantum information theory. Although the first famous paradox connected with
entanglement \cite{EPR35} was presented in the context of observables with
continuous values, the specific entanglement properties of such systems
\cite{RevModPhys.77.513,PhysRevLett.80.869,PhysRevA.61.022309,Simon00} are
less explored than those with discrete, e.g. spin states.

Recent studies of continuous variable quantum systems focus on the emergence
of bipartite entanglement in a scattering event of two interacting
distinguishable quantum particles, which have no initial correlations
\cite{Sch04,TK05, WLC05, WLC06, BF07, L04}. Since this is a fundamental
process in quantum physics, it is important to explore how does it generate
quantum entanglement. Some general features of this process were identified in
\cite{HarshmanPhysRevA2008,HarshmanJPhysA2008}. Refs.
\cite{FreybergerPRA2002,WLC05,BF07} considered specific models on the
scattering of ultracold atoms trapped in a harmonic potential well. Important
results on the entanglement of colliding particles, modelled by Gaussian wave
packets and interacting with different finite range potentials, were published
in \cite{Sch98,Sch04,L04,TK05,SchJ06,HarshmanJPhysA2008}.

In the present work we consider the explicit time dependence of entanglement
in a quantum mechanical model of a collision process which itself creates the
entanglement during the interaction between two particles that were
independent in the beginning. To be specific, we assume a non-relativistic one
dimensional motion with an attractive or repulsive delta potential between the
particles. The evolution of the process is described by the explicit solution
of the time dependent Schrödinger equation. Although this problem has
been considered previously, here we derive analytic expressions for the post
collision behaviour that incorporate the spread of the individual wave
packets, by using the exact propagator for the delta potential
\cite{Elberfeld_Kleber_1988,Blinder88}. To quantify the entanglement we use the Neumann entropy
\cite{N55}, and we present how it is built up during the collision. The
asymptotic value of the entropy is then obtained from our analytic expression
of the long time limit of the time dependent two-particle wave function. This
study may find application e.g. in the experimental analysis of collision and
recollision of atomic fragments following a laser induced dissociation \cite{Vrakking_etal_2011}.

\section{Interaction of two particles via a delta potential}

The Hamiltonian of the system is written in terms of the position and momentum
operators of the particles as
\begin{equation}
H_{12}=\frac{P_{1}^{2}}{2m_{1}}+\frac{P_{2}^{2}}{2m_{2}}-V_{0}\delta\left(
X_{1}-X_{2}\right)  . \label{eq7}%
\end{equation}
For an attractive (repulsive) interaction we have here $V_{0}>0,$ $(V_{0}<0)$.
We introduce, as usual for two-body problems, the operators:%

\begin{eqnarray}
X_{0}  & = \alpha_{1}X_{1}+\alpha_{2}X_{2}, \qquad  P_{0}  &  =P_{1}+P_{2}%
\label{relcent}\\
X &  = X_{1}-X_{2}, \qquad\qquad  P  &  =\alpha_{2}P_{1}-\alpha_{1}P_{2}%
\end{eqnarray}
resulting in a sum of two independent Hamiltonians corresponding to the center
of mass motion and the relative motion: \
\begin{equation}
H_{12}=H_{0}+H,\qquad H_{0}=\frac{P_{0}^{2}}{2m},\qquad H=\frac{P^{2}}{2\mu
}-V_{0}\delta(X).
\end{equation}
Here $m=m_{1}+m_{2}$ is the total mass of the particles, $\alpha_{i}=m_{i}/m$
$\ $and$\ \mu=m_{1}m_{2}/(m_{1}+m_{2})$\ is the reduced mass of the system. In
the center of mass reference frame the expectation value of $P_{0}$, the total
momentum of the particles, is zero, and the natural coordinate system is the
one which has its origin in the expectation value of the center of mass
operator, $X_{0}$. Then $\left\langle P_{0}\right\rangle =0$ and $\left\langle
X_{0}\right\rangle =0$ for all times. We shall proceed by using coordinate
wave functions and assume that initially the particles are described by a
product of normalized Gaussians
\begin{eqnarray}
\Psi\left(  x_{1},x_{2},t=0\right)   & = & \varphi_{1}(x_{1})\varphi_{2}%
(x_{2}) \label{init}\\
&  = & \frac{(\alpha_{1}\alpha_{2})^{1/4}}{\sigma\sqrt{\pi}}e^{-\alpha_{1}\left(
x_{1}+a_{1}\right)  ^{2}/2\sigma^{2}}e^{ik_{1}x_{1}}e^{-\alpha_{2}\left(
x_{2}-a_{2}\right)  ^{2}/2\sigma^{2}}e^{ik_{2}x_{2}}%
\end{eqnarray}
localized at distant points: around $-a_{1}=-\alpha_{2}a$ and $a_{2}%
=\alpha_{1}a,$ as required by $\left\langle X_{0}\right\rangle =0.$ From
$\left\langle P_{0}\right\rangle =0$ we also have $-k_{2}=k_{1}=:q$. In terms
of the center of mass and relative coordinates, $x_{0}$ and $x$ this wave
function takes the form:%
\begin{equation}
\Psi\left(  x_{1},x_{2},t=0\right)  =\Phi\left(  x_{0},x,t=0\right)
=\varphi_{c}\left(  x_{0}\right)  \varphi_{r}\left(  x\right) ,
\end{equation}
where
\begin{eqnarray}
\varphi_{c}\left(  x_{0}\right)   &  = & \sigma^{-1/2}\pi^{-1/4}e^{-x_{0}%
^{2}/2\sigma^{2}},\qquad\label{fi0c}\\
\varphi_{r}\left(  x\right)   &  = & (\alpha/\sigma)^{1/2}\pi^{-1/4}%
e^{-\alpha^{2}\left(  x+a\right)  ^{2}/2\sigma^{2}}e^{iqx}\qquad\label{fir0}%
\end{eqnarray}
are normalized functions of $x_{0}$ and $x,$ respectively, $a=a_{1}+a_{2}$ is
the mean value of the distance between the particles, and $\alpha^{2}%
=\alpha_{1}\alpha_{2}=\mu/m$.

The separability of the wave function in terms of the coordinates 
$x$ and $x_{0}$ is due to the specific choice of the initial
wave function \cite{Sch98,Sch04,SchJ06,HarshmanJPhysA2008,HarshmanPhysRevA2008}, 
where the variances of
the positions of the individual particles $(\Delta X_{i})^{2}$ obey
$m_{1}(\Delta X_{1})^{2}=m_{2}(\Delta X_{2})^{2}$. In the more
general case, one has a double sum of products of arbitrary basis functions in
the new variables, which could still be transformed into a single sum in the
Schmidt bases of the respective spaces (see Eq. (\ref{SD}) below). As implied
by the linearity of the Schrödinger equation, the time evolution of the
initial state could then be obtained by solving the problem for each term in
the sum.

The time evolution of the wave function in the coordinates $x_{0}$ and $x$ are
determined by $H_{0}$ and $H$ independently, and they can be given by the
respective propagators. For the free motion of the center of mass this is well
known:%
\begin{equation}
K_{m}^{0}(x_{0},y_{0},t)=\left(  \frac{m}{2\pi i\hbar t}\right)  ^{1/2}%
\exp\left[  im(x_{0}-y_{0})^{2}/2\hbar t\right]  \label{K0}%
\end{equation}
which yields the usual spreading Gaussian wave packet according to:%
\begin{eqnarray}
\Phi_{c}\left(  x_{0},t\right)   &  =\int K_{m}^{0}(x_{0},y_{0},t)\varphi
_{c}(y_{0})dy_{0}=\nonumber\\
&  =N_{t}\exp\left[  -x_{0}^{2}/2\sigma_{t}^{2}\right]
\end{eqnarray}
where $N_{t}=\pi^{-1/4}(\sigma+i\hbar t/m\sigma)^{-1/2}$ and $\sigma_{t}%
^{2}=\sigma^{2}+i\hbar t/m.$ The propagator for the delta potential
Hamiltonian is more complicated, but still can be obtained in a closed form.
For the attractive case ($V_{0}>0$) the propagator has been derived in
\cite{Blinder88}, while it turns out that the result is valid for both signs
of the potential and is given by:
\begin{equation}
K(x,y,t)=K_{\mu}^{0}(x,y,t)+\frac{g}{2}e^{-\frac{(|x|+|y|)^{2}}{4\beta}%
}e^{u^{2}}\mbox{erfc}\left(  u\right)  ,\label{deltprop}%
\end{equation}
where
\begin{equation}
g=\frac{\mu V_{0}}{\hbar^{2}},\qquad\beta=\frac{i\hbar t}{2\mu},\qquad
u=\frac{|x|+|y|}{\sqrt{2i\hbar t/\mu}}-g\sqrt{i\hbar t/2\mu}%
\end{equation}
The time dependence of the relative wave function can now be given as:
\begin{equation}
\Phi_{r}\left(  x,t\right)  =\int K(x,y,t)\varphi_{r}(y)dy\label{FIxt}%
\end{equation}
\qquad It is not possible to determine $\Phi_{r}\left(  x,t\right)  $ in a
closed form given its initial value by (\ref{fir0}), we have to rely on
numerical integration, but a very good approximate formula, valid for large
times, will be given below.

In order to consider the entanglement of the particles one makes the
substitution corresponding to (\ref{relcent}) and obtains a function of
$x_{1}$ and $x_{2}$:
\begin{equation}
\Psi\left(  x_{1},x_{2},t\right)  =\Phi_{c}\left(  x_{0},t\right)  \Phi
_{r}\left(  x,t\right)  =\Phi_{c}(\alpha_{1}x_{1}+\alpha_{2}x_{2},t)\Phi
_{r}(x_{1}-x_{2},t). \label{Psit}%
\end{equation}
which is not a separable state in the original coordinates $x_{1}$ and $x_{2}$
any more. The above mentioned approximation for $\Phi_{r}\left(  x,t\right)  $
will enable us to treat $\Psi\left(  x_{1},x_{2},t\right)  $ analytically, and
consider explicitly the entanglement involved in it.

\section{Reduced density operator and entanglement}

In order to quantify entanglement in the state in (\ref{Psit}) one uses a
measure that characterizes how much an actual two-particle wave function is
different from a single product of two one-particle wave functions. In the
context of quantum mechanics this was formulated first by J. Neumann
\cite{N55}, based on the Schmidt decomposition \cite{Sch07} theorem. It states
that for a square integrable function $\Psi(x_{1},x_{2},t)$ of two variables,
there exist a set of functions $\phi_{k}(x_{1},t)$ and $\psi_{k}(x_{2},t)$
which both form an orthonormal (but not necessarily complete) set in their
respective Hilbert spaces, such that $\Psi$ can be written as a single sum of
their products:\
\begin{equation}
\Psi(x_{1},x_{2},t)=\sum\limits_{k}\sqrt{p_{k}(t)}\phi_{k}(x_{1},t)\psi
_{k}(x_{2},t), \label{SD}%
\end{equation}
The actual values of the $p_{k}$-s are the simultaneous eigenvalues of the
reduced density operators $\hat{\varrho}_{1}$ and $\hat{\varrho}_{2}$
describing either of the two subsystems defined by the hermitian kernel: \
\begin{equation}
\varrho_{1}(x_{1}^{\prime},x_{1},t)=\int\Psi^{\ast}(x_{1}^{\prime}%
,x_{2},t)\Psi(x_{1},x_{2},t)dx_{2} \label{red1}%
\end{equation}
for system 1 and similarly for system 2. As it is shown in \cite{N55,Sch07} in
more detail, $\hat{\varrho}_{1}$ and $\hat{\varrho}_{2}$ have a complete set
of square integrable eigenfunctions forming a basis, $\phi_{k}(x_{1})$,
$\psi_{k}(x_{2})$ respectively, such that the corresponding eigenvalues
$p_{k}$ are identical. The $p_{k}$-s are nonnegative and form a discrete set,
the sum of which is equal to unity. This also implies that the multiplicity of
a positive eigenvalue must be finite.

For quantifying entanglement it is natural to use the measure of randomness of
the discrete probability distribution given by the $p_{k}$-s in the Schmidt
sum. Statistical physics tells us that this is best characterized by
$S=-\sum\limits_{k}p_{k}\ln p_{k}$, which is just the Neumann entropy
belonging to the reduced density operator for each of the particles%
\begin{equation}
S(t)=-\mbox{Tr}\left[  \hat{\rho}_{1}(t)\ln\hat{\rho}_{1}(t)\right]
=-\mbox{Tr}\left[  \hat{\rho}_{2}(t)\ln\hat{\rho}_{2}(t)\right]  \label{S}%
\end{equation}

\begin{figure}[h]
\includegraphics[width=6in]{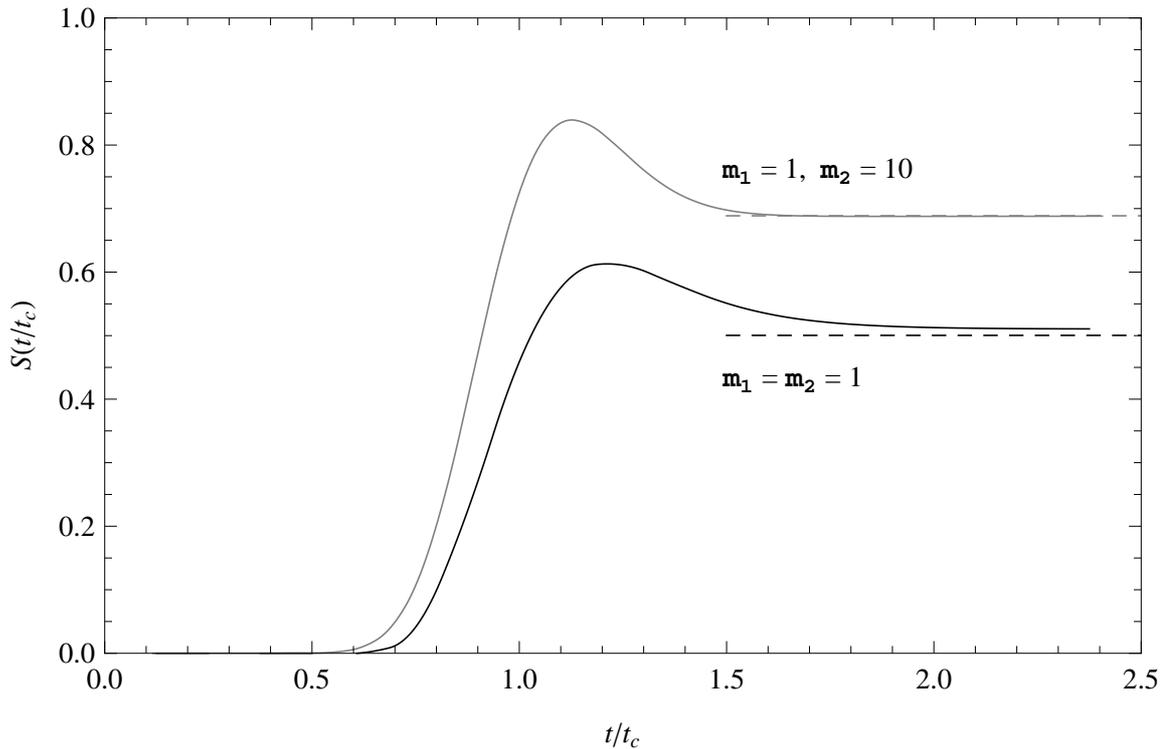}
\caption{Time dependence of the quantum entanglement during the collision
process. We plot the Neumann entropy (\ref{S}), computed numerically, vs. time
measured in units of the classical collision time $t_{c}=ma/\hbar q$. The
black solid curve is for $m_{1}=m_{2}=1$, while the gray solid curve is for
$m_{1}=1$, $m_{2}=10$, as indicated in the figure. The dashed lines mark the
asymptotic values of $S$, given by formula (\ref{asymptS}). We use atomic
units, the parameters are: $V_{0}=-5, \, a=10, \, q=5,\, \sigma=1/2$. }%
\label{fig:S(t)}%
\end{figure}

In order to calculate $S$ one has to find the nonzero eigenvalues $p_{k}$,
that shall be time dependent during the collision. Fig. \ref{fig:S(t)} shows
numerical results about the time dependence of the quantum entanglement during
the collision process, using atomic units.

\section{Approximate propagator and entropy for long times}

The time evolution of the relative wave function, i.e. the second factor in
(\ref{FIxt}) cannot be given in a closed form. We find an approximate
analytical formula for the relative wave function and entropy using certain
approximations for the propagator given by (\ref{deltprop}).

The assumption that initially the particles are localized \ in a large
distance from each other means that $\varphi_{r}(y)$ is different from zero
only around $-a\ll0$, therefore the contribution to the integral in
(\ref{FIxt}) for $y>0$ can be neglected, and $|y|$ can be replaced by $-y$ in
the propagator (\ref{deltprop}).\ Setting $v=\hbar q/m,$ which is the mean
velocity of the relative wave packet, we can consider the asymptotic behaviour
of the system for times much larger than $t_{c}=a/v$, which is the time
instant of the corresponding classical collision. We can use then the
asymptotic approximation \cite{AS65}: $\exp(u^{2})\mbox{erfc}(u) = (\sqrt{\pi
}\,{u})^{-1}(1-{u^{-2}}/2+O({u^{-4}}))$, valid for large values of $u$.
Keeping only the first term here we have then:
\begin{eqnarray}
K = K_{0}^{\mu}+K_{1}^{\mu} =  &  (4\pi\beta)^{-1/2}\exp\left[  -
\frac{(x-y)^{2}}{4\beta}\right]  +
\nonumber\\
&  \frac{g\sqrt{\beta}}{\sqrt{\pi}\left(  |x|-y-2g\beta\right)  }\exp\left[
-\frac{(|x|-y)^{2}}{4\beta}\right]  . \label{Kapr1}%
\end{eqnarray}

The reduced (one-particle) density matrix in the form of Eq.\ (\ref{red1})
cannot be calculated from the propagator (\ref{Kapr1}) analytically.
Therefore, we simplify it further by replacing the $x$ and $y$ variables of
the propagator with the classical initial and final coordinate values, $y=-a$
and $|x|=|-a+\hbar q t/\mu|$, in the prefactor of the exponential of $K_{1}$,
but keeping the position dependence in the rapidly oscillating phase factor.
This approximation is similar to the usual one in scattering theory, and
yields the propagator
\begin{eqnarray}
K(x,y,t) & =  &  (4\pi\beta)^{-1/2}\left(  \exp\left[  -\frac{(x-y)^{2}}{4\beta
}\right]  \right. \nonumber\\
& &  \left.  +\left(  \frac{|-a+\hbar qt/\mu|+a}{it\hbar g/\mu}-1\right)  ^{-1}
\exp\left[  -\frac{(|x|-y)^{2}}{4\beta}\right]  \right)  .
\end{eqnarray}
Due to the presence of $|x|$ in the exponential of the second term, the
integral (\ref{FIxt}) with the function $\varphi_{r}(y)$ from (\ref{fir0})
will split into two distinct Gaussians: one centered around $x=-a+vt$, this
corresponds to the forward scattered wave, while the other around $a-vt$
\ yielding the reflected wave. In this way we get the following asymptotic
form of the relative propagator:%

\begin{equation}
K(x,y,t)=\sqrt{\frac{\mu}{2\pi i\hbar t}} \left(  c_{+} \, \exp\left[
\frac{i\mu(x-y)^{2}}{2\hbar t} \right]  + c_{-} \, \exp\left[  \frac
{i\mu(x+y)^{2}}{2\hbar t} \right]  \right)  \label{Kapr2}%
\end{equation}
where $c_{-}=\left(  \frac{|-a+vt|+a}{i\hbar tg/\mu}-1\right)  ^{-1}$,
$c_{+}=1+c_{-}$. It is easy to check that for times larger than $v/a$ these
amplitudes coincide with the plane wave transmission and reflection
coefficients for a delta potential with wave number $q,$ which is the mean
value of $k$ in the initial relative state:%
\begin{equation}
\lim_{t \rightarrow\infty} c_{+}=T(q):=q/(q-ig),\qquad\lim_{t \rightarrow
\infty} c_{-}=R(q):=ig/(q-ig)
\end{equation}

The great advantage of the approximate form in (\ref{Kapr2}) is that it allows
one to proceed entirely analytically and determine the total wave function, as
well as the final value of the entanglement entropy in the system in a closed
form. This is due to the emergence of Gaussian type integrals in (\ref{FIxt}),
which leads us to the approximate relative wave function%

\begin{equation}
\tilde{\Phi}_{r}(x,t)=T(q)\phi_{+}(x,t)+R(q)\phi_{-}(x,t)
\end{equation}
where
\begin{equation}
\phi_{\pm}(x,t)=N_{t}\sqrt{\alpha}\exp\left[  -\frac{\alpha^{2}(\pm
x+a-vt)^{2}}{2(\sigma^{2}+i\hbar t/m)}\right]  .
\end{equation}

The total two-particle wave function is now obtained with Eq. (\ref{Psit}),
and after some algebra we obtain the result
\begin{equation}
\Psi(x_{1},x_{2},t)=\left[  T(q)\Phi_{1}(x_{1},t)\Phi_{2}(x_{2},t)+R(q)\Phi
_{1}(-x_{1},t)\Phi_{2}(-x_{2},t)\right]  \label{Psias}%
\end{equation}
with
\begin{equation}
\Phi_{j}(\xi,t)=N_{t}\alpha_{j}^{1/4}\exp\left[  -\frac{\alpha_{j}(\xi
-u_{1}t+a_{j})^{2}+i\gamma_{j}(\xi,t)}{\sigma_{t}^{2}}\right]  \label{Phias}%
\end{equation}
and
\begin{equation}
\gamma_{j}(\xi,t)=\frac{\hbar t}{m\sigma^{2}}[\alpha_{j}(a_{j}+\xi)^{2}%
-\alpha^{2}a^{2}/2]+\sigma^{2}q[2(a_{j}+\xi)-vt/2]
\end{equation}
Here $a_{1} = \alpha_{2} a$, $a_{2} = \alpha_{1} a$, the $\sigma_{t}%
^{2}/\alpha_{j}=(\sigma^{2}+\hbar^{2}t^{2}/\sigma^{2}m^{2})/\alpha_{j}$
describes the spreading of the respective wave packets, and $u_{j}=\hbar
q/m_{j}$ are the velocities of the particles in the corresponding classical problem.

\begin{figure}[htp]
\includegraphics[width=6in]{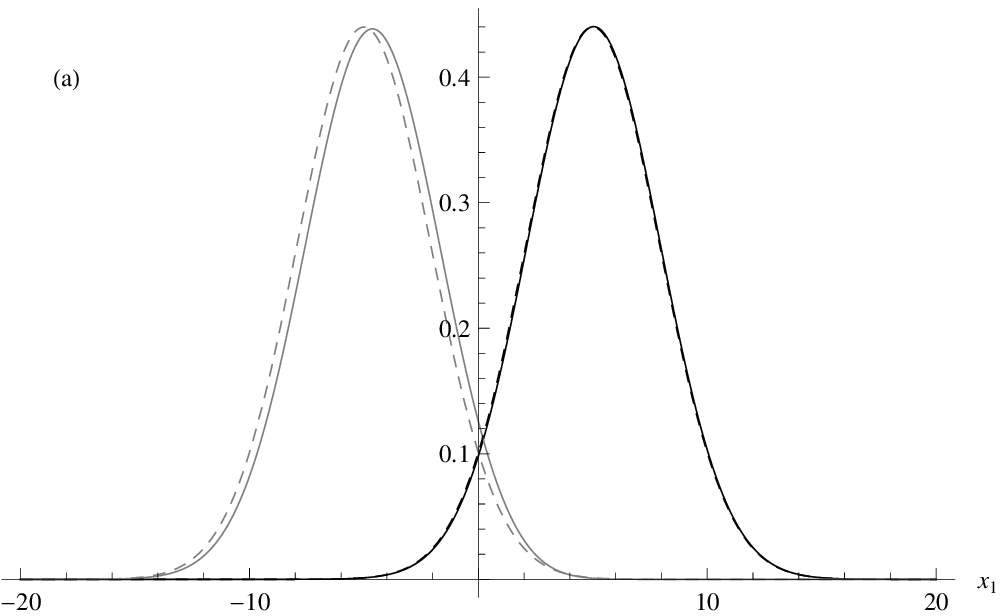}
\includegraphics[width=6in]{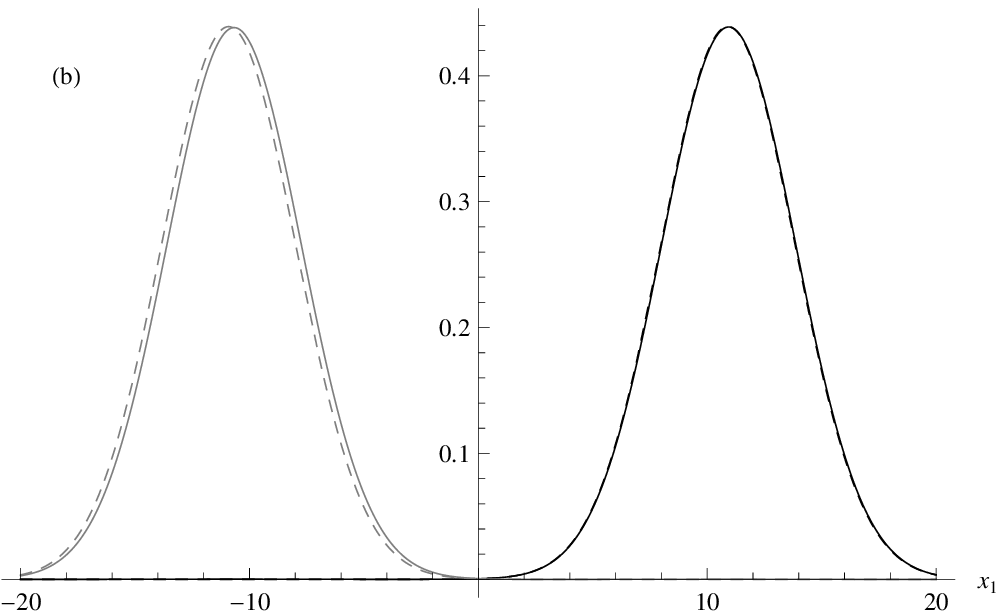}
\caption{Comparison of the asymptotic analytical results of the Schmidt
decomposition, Eq. (\ref{Psias}), with the numerical method, for the case of
(a) $m_{1}=m_{2}=1$, at $t=2 \, t_{c}$, and (b) $m_{1}=1$, $m_{2}=10$, at
$t=2\, t_{c}$. We plot the absolute value of the two eigenstates of the one-particle density matrix
having the two largest eigenvalues, obtained numerically, with solid black (on
the right side) and solid gray (on the left side) curves. The dashed black and
dashed gray curves are the absolute values of $\Phi_{1}(x_{1},t)$ and
$\Phi_{1}(-x_{1},t)$ functions, respectively, see Eq. (\ref{Phias}). We use
atomic units, the parameters are: $V_{0}=-5, \, a=10, \, q=5,\, \sigma=1/2$. }%
\label{fig:Schmidt}%
\end{figure}

Fig. \ref{fig:Schmidt} shows $\Phi_{1}(x_{1},t)$ and $\Phi_{1}(-x_{1},t)$ in
comparison with the two eigenstates of the computed one-particle density
matrix, having the largest eigenvalues.

The expression (\ref{Psias}) is the required approximation of the Schmidt
decomposition, consisting of the two terms. The asymptotic value of the
entropy of this entangled state is then
\begin{eqnarray}
S  &  =-|T(q)|^{2}\ln|T(q)|^{2}-|R(q)|^{2}\ln|R(q)|^{2}=\nonumber\\
&  =-(\frac{q^{2}}{q^{2}+g^{2}}\ln\frac{q^{2}}{q^{2}+g^{2}}+\frac{g^{2}}%
{q^{2}+g^{2}}\ln\frac{g^{2}}{q^{2}+g^{2}}) \label{asymptS}%
\end{eqnarray}
The entanglement will be maximal for $q=|g|$, i.e. for center of mass momenta
with $-p_{2}=p_{1}=\mu|V_{0}|/\hbar$, and has the value $\ln2$. This
explains why we have a larger entanglement for $m_{2}=10m_{1}$, when
this condition is almost satisfied.

\section{Conclusions}

We have followed the emergence of the entanglement of two colliding particles
that are independent initially and thus their wave function is a product
state. In the case of the delta interaction potential and Gaussian initial
states we have found an approximate analytic expression for the final
entangled state, which is in good agreement with numerical results based on
the exact propagator. Let us note that the wave function $\Psi( x_{1},x_{2},t
) $ is a kind of an EPR\ state \cite{EPR35} in its original sense, i.e. in the
coordinate space of two entangled particles. In \cite{EPR35} however the state
considered is given by a highly singular delta function, while here
$\Psi\left(  x_{1},x_{2},t\right)  $ is square integrable during the whole
process and has an especially simple asymptotic form.

\section*{Acknowledgments}

This research has been granted by the Hungarian Scientific Research Fund OTKA
under Contracts No. T81364, and by the \textquotedblleft
TAMOP-4.2.1/B-09/1/KONV-2010-0005 project: Creating the Center of Excellence
at the University of Szeged\textquotedblright\ supported by the EU and the
European Regional Development Fund.


\section*{References}

\begin{thebibliography}{22}
\expandafter\ifx\csname natexlab\endcsname\relax\def\natexlab#1{#1}\fi
\expandafter\ifx\csname bibnamefont\endcsname\relax
  \def\bibnamefont#1{#1}\fi
\expandafter\ifx\csname bibfnamefont\endcsname\relax
  \def\bibfnamefont#1{#1}\fi
\expandafter\ifx\csname citenamefont\endcsname\relax
  \def\citenamefont#1{#1}\fi
\expandafter\ifx\csname url\endcsname\relax
  \def\url#1{\texttt{#1}}\fi
\expandafter\ifx\csname urlprefix\endcsname\relax\def\urlprefix{URL }\fi
\providecommand{\bibinfo}[2]{#2}
\providecommand{\eprint}[2][]{\url{#2}}

\bibitem[{\citenamefont{Einstein et~al.}(1935)\citenamefont{Einstein, Podolsky,
  and Rosen}}]{EPR35}
\bibinfo{author}{\bibfnamefont{A.}~\bibnamefont{Einstein}},
  \bibinfo{author}{\bibfnamefont{B.}~\bibnamefont{Podolsky}}, \bibnamefont{and}
  \bibinfo{author}{\bibfnamefont{N.}~\bibnamefont{Rosen}},
  \bibinfo{journal}{Phys. Rev.} \textbf{\bibinfo{volume}{47}},
  \bibinfo{pages}{777} (\bibinfo{year}{1935}).

\bibitem[{\citenamefont{Braunstein and van Loock}(2005)}]{RevModPhys.77.513}
\bibinfo{author}{\bibfnamefont{S.}~\bibnamefont{Braunstein}} \bibnamefont{and}
  \bibinfo{author}{\bibfnamefont{P.}~\bibnamefont{van Loock}},
  \bibinfo{journal}{Rev. Mod. Phys.} \textbf{\bibinfo{volume}{77}},
  \bibinfo{pages}{513} (\bibinfo{year}{2005}).

\bibitem[{\citenamefont{Braunstein and Kimble}(1998)}]{PhysRevLett.80.869}
\bibinfo{author}{\bibfnamefont{S.}~\bibnamefont{Braunstein}} \bibnamefont{and}
  \bibinfo{author}{\bibfnamefont{H.}~\bibnamefont{Kimble}},
  \bibinfo{journal}{Phys. Rev. Lett.} \textbf{\bibinfo{volume}{80}},
  \bibinfo{pages}{869} (\bibinfo{year}{1998}).

\bibitem[{\citenamefont{Hillery}(2000)}]{PhysRevA.61.022309}
\bibinfo{author}{\bibfnamefont{M.}~\bibnamefont{Hillery}},
  \bibinfo{journal}{Phys. Rev. A} \textbf{\bibinfo{volume}{61}},
  \bibinfo{pages}{022309} (\bibinfo{year}{2000}).

\bibitem[{\citenamefont{Simon}(2000)}]{Simon00}
\bibinfo{author}{\bibfnamefont{R.}~\bibnamefont{Simon}},
  \bibinfo{journal}{Phys. Rev. Lett.} \textbf{\bibinfo{volume}{84}},
  \bibinfo{pages}{2726} (\bibinfo{year}{2000}).

\bibitem[{\citenamefont{Schulman}(2004)}]{Sch04}
\bibinfo{author}{\bibfnamefont{L.}~\bibnamefont{Schulman}},
  \bibinfo{journal}{Phys. Rev. Lett.} \textbf{\bibinfo{volume}{92}},
  \bibinfo{pages}{210404} (\bibinfo{year}{2004}).

\bibitem[{\citenamefont{Tal and Kurizki}(2005)}]{TK05}
\bibinfo{author}{\bibfnamefont{A.}~\bibnamefont{Tal}} \bibnamefont{and}
  \bibinfo{author}{\bibfnamefont{G.}~\bibnamefont{Kurizki}},
  \bibinfo{journal}{Phys. Rev. Lett.} \textbf{\bibinfo{volume}{94}},
  \bibinfo{pages}{160503} (\bibinfo{year}{2005}).

\bibitem[{\citenamefont{Wang et~al.}(2005)\citenamefont{Wang, Law, and
  Chu}}]{WLC05}
\bibinfo{author}{\bibfnamefont{J.}~\bibnamefont{Wang}},
  \bibinfo{author}{\bibfnamefont{C.}~\bibnamefont{Law}}, \bibnamefont{and}
  \bibinfo{author}{\bibfnamefont{M.}~\bibnamefont{Chu}},
  \bibinfo{journal}{Phys. Rev. A} \textbf{\bibinfo{volume}{72}},
  \bibinfo{pages}{022346} (\bibinfo{year}{2005}).

\bibitem[{\citenamefont{Wang et~al.}(2006)\citenamefont{Wang, Law, and
  Chu}}]{WLC06}
\bibinfo{author}{\bibfnamefont{J.}~\bibnamefont{Wang}},
  \bibinfo{author}{\bibfnamefont{C.}~\bibnamefont{Law}}, \bibnamefont{and}
  \bibinfo{author}{\bibfnamefont{M.}~\bibnamefont{Chu}},
  \bibinfo{journal}{Phys. Rev. A} \textbf{\bibinfo{volume}{73}},
  \bibinfo{pages}{034302} (\bibinfo{year}{2006}).

\bibitem[{\citenamefont{Busshardt and Freyberger}(2007)}]{BF07}
\bibinfo{author}{\bibfnamefont{M.}~\bibnamefont{Busshardt}} \bibnamefont{and}
  \bibinfo{author}{\bibfnamefont{M.}~\bibnamefont{Freyberger}},
  \bibinfo{journal}{Phys. Rev. A} \textbf{\bibinfo{volume}{75}},
  \bibinfo{pages}{052101} (\bibinfo{year}{2007}).

\bibitem[{\citenamefont{Law}(2004)}]{L04}
\bibinfo{author}{\bibfnamefont{C.}~\bibnamefont{Law}}, \bibinfo{journal}{Phys.
  Rev. A} \textbf{\bibinfo{volume}{70}}, \bibinfo{pages}{062311}
  (\bibinfo{year}{2004}).

\bibitem[{\citenamefont{Harshman and Hutton}(2008)}]{HarshmanPhysRevA2008}
\bibinfo{author}{\bibfnamefont{N.~L.} \bibnamefont{Harshman}} \bibnamefont{and}
  \bibinfo{author}{\bibfnamefont{G.}~\bibnamefont{Hutton}},
  \bibinfo{journal}{Phys. Rev. A} \textbf{\bibinfo{volume}{77}},
  \bibinfo{pages}{042310} (\bibinfo{year}{2008}).

\bibitem[{\citenamefont{Harshman and Singh}(2008)}]{HarshmanJPhysA2008}
\bibinfo{author}{\bibfnamefont{N.~L.} \bibnamefont{Harshman}} \bibnamefont{and}
  \bibinfo{author}{\bibfnamefont{P.}~\bibnamefont{Singh}}, \bibinfo{journal}{J.
  Phys. A-Math. Theor.} \textbf{\bibinfo{volume}{41}}, \bibinfo{pages}{155304}
  (\bibinfo{year}{2008}).

\bibitem[{\citenamefont{Mack and Freyberger}(2002)}]{FreybergerPRA2002}
\bibinfo{author}{\bibfnamefont{H.}~\bibnamefont{Mack}} \bibnamefont{and}
  \bibinfo{author}{\bibfnamefont{M.}~\bibnamefont{Freyberger}},
  \bibinfo{journal}{Phys. Rev. A} \textbf{\bibinfo{volume}{66}},
  \bibinfo{pages}{042113} (\bibinfo{year}{2002}).

\bibitem[{\citenamefont{Schulman}(1998)}]{Sch98}
\bibinfo{author}{\bibfnamefont{L.~S.} \bibnamefont{Schulman}},
  \bibinfo{journal}{Phys. Rev. A} \textbf{\bibinfo{volume}{57}},
  \bibinfo{pages}{840} (\bibinfo{year}{1998}).

\bibitem[{\citenamefont{Schmuser and Janzing}(2006)}]{SchJ06}
\bibinfo{author}{\bibfnamefont{F.}~\bibnamefont{Schmuser}} \bibnamefont{and}
  \bibinfo{author}{\bibfnamefont{D.}~\bibnamefont{Janzing}},
  \bibinfo{journal}{Phys. Rev. A} \textbf{\bibinfo{volume}{73}},
  \bibinfo{pages}{052313} (\bibinfo{year}{2006}).

\bibitem[{\citenamefont{Elberfeld and Kleber}(1988)}]{Elberfeld_Kleber_1988}
\bibinfo{author}{\bibfnamefont{W.}~\bibnamefont{Elberfeld}} \bibnamefont{and}
  \bibinfo{author}{\bibfnamefont{M.}~\bibnamefont{Kleber}},
  \bibinfo{journal}{Am. J. Phys.} \textbf{\bibinfo{volume}{56}},
  \bibinfo{pages}{154} (\bibinfo{year}{1988}).

\bibitem[{\citenamefont{Blinder}(1988)}]{Blinder88}
\bibinfo{author}{\bibfnamefont{S.~M.} \bibnamefont{Blinder}},
  \bibinfo{journal}{Phys. Rev. A} \textbf{\bibinfo{volume}{37}},
  \bibinfo{pages}{973} (\bibinfo{year}{1988}).

\bibitem[{\citenamefont{Neumann}(1955)}]{N55}
\bibinfo{author}{\bibfnamefont{J.}~\bibnamefont{Neumann}},
  \emph{\bibinfo{title}{Mathematical Foundations of Quantum Mechanics}}
  (\bibinfo{publisher}{Princeton University Press},
  \bibinfo{address}{Princeton}, \bibinfo{year}{1955}).

\bibitem[{\citenamefont{Kelkensberg et~al.}(2011)}]{Vrakking_etal_2011}
\bibinfo{author}{\bibfnamefont{F.}~\bibnamefont{Kelkensberg}}
  \bibnamefont{et~al.}, \bibinfo{journal}{Phys. Rev. Lett.}
  \textbf{\bibinfo{volume}{107}}, \bibinfo{pages}{043002}
  (\bibinfo{year}{2011}).

\bibitem[{\citenamefont{Schmidt}(1907)}]{Sch07}
\bibinfo{author}{\bibfnamefont{E.}~\bibnamefont{Schmidt}},
  \bibinfo{journal}{Math. Ann.} \textbf{\bibinfo{volume}{63}},
  \bibinfo{pages}{43} (\bibinfo{year}{1907}).

\bibitem[{\citenamefont{Abramowitz and Stegun}(1965)}]{AS65}
\bibinfo{editor}{\bibfnamefont{M.}~\bibnamefont{Abramowitz}} \bibnamefont{and}
  \bibinfo{editor}{\bibfnamefont{I.}~\bibnamefont{Stegun}}, eds.,
  \emph{\bibinfo{title}{Handbook of mathematical functions}}
  (\bibinfo{publisher}{Dover Publications}, \bibinfo{address}{New York},
  \bibinfo{year}{1965}).

\end{thebibliography}

\end{document}